\documentclass[prb,aps,twocolumn,showpacs,showkeys]{revtex4}
\begin{document}

\title{Universal Landauer conductance in chiral symmetric $2d$ systems}
\author{Daniel G.\ Barci}

\affiliation{Departamento de F{\'\i}sica Te\'orica,
Universidade do Estado do Rio de Janeiro, Rua S\~ao Francisco Xavier 524, 20550-013,  Rio de Janeiro, RJ, Brazil.}
\altaffiliation{Research Associate of the Abdus Salam International Centre for Theoretical Physics}

\author{Luis E. Oxman}
\affiliation{Instituto de F\'{\i}sica, Universidade Federal Fluminense\\
Av. Litor\^anea S/N, Boa Viagem, Niter\'oi, RJ 24210-340, Brazil.}
\date{\today}

\begin{abstract}
We study transport properties of an arbitrarily shaped ultraclean graphene sheet, adiabatically connected to leads,
composed by the same material. 
If the localized interactions do not destroy chiral symmetry, we show that the conductance is quantized, since it is dominated by the quasi one-dimensional leads. As an example, we show that smooth structural deformations of the graphene plane do not modify the conductance quantization.

\end{abstract}

\pacs{73.23.Ad, 72.10.Bg, 71.10.Pm, 71.10.Fd}
%73.23.Ad  Ballistic transport
%72.10.Bg  General formulation of transport theory
%71.10.Pm  Fermions in reduced dimensions (anyons, composite fermions,Luttinger liquid, etc.)
%71.10.Fd  Lattice fermion models (Hubbard model, etc.)
%72.15.Lh  Relaxation times and mean free paths

\keywords{graphene, bosonization, chiral symmetry}
\maketitle

\section{Introduction}

In recent  years an explosive amount of attention has been paid to the novel material graphene\cite{CastroNetoReview}, a $2d$ carbon sheet.
One of the important reasons for this great interest is the peculiar behavior of the associated transport properties. 

Essentially, graphene is a semi-metal with a gap over almost all the $2d$ Brillowin zone, except for two symmetrical points where the low energy excitations are gapless fermions with a linear dispersion relation\cite{first direct}.
The low energy description of graphene can be done in terms of four-component massless Dirac fermions. In this scenario, curious behaviors such as quantized conductivity\cite{qcon1,qcon2,qcon3,qcon4,qcon5,qcon6}, unconventional integer quantum Hall effect\cite{qcon2,unc1,unc2}, and the Klein paradox\cite{KP} (unimpeded penetration through high and wide potential barriers), can be discussed.

With regard to conductance, although current experiments observe a dissipative (Ohmic) regime\cite{Exp1}, from a theoretical point of view, the study of transport properties of charged $2d$ massless Dirac modes is a fascinating problem, far away from its complete understanding\cite{Theory1}. 

In ref. \onlinecite{CN2006}, for a thin ultraclean graphene strip, a quantized conductance related with the zero modes at the edge of the strip was found; similarly to what happens in nanotubes, where evidence of conductance quantization was recently reported\cite{NanotubesNature}. The value of the conductance depends on whether the edges are cut with zig zag or armchair geometry. In that work, a one particle approximation was considered,  disregarding the effect of any kind of impurities or interactions. 

In this work, we claim that this behavior is not only correct, but it can be extended to more general situations. The conductance quantization is preserved if, in a localized region, the system suffers an adiabatic change with respect to the situation at the strip, preserving  chiral symmetry.

We recall that, in planar models involving a four-component spinor $\psi$, two independent matrices $\gamma^3$,
$\gamma^4$ that anticommute with the Dirac matrices can be defined, which play a similar role to the usual $\gamma^5$, in one and three-dimensional Dirac theories.

At the classical level, a planar system of massless Dirac fermions, including (total) charge density and current interactions, are symmetric under the continuous chiral transformations, generated by $\gamma^3$, $\gamma^4$.
Then, as examples of adiabatic change, we can consider an adiabatic widening of the $2d$ sample, or even effective current interactions coming, for instance, from a smooth deformation of the graphene plane\cite{Kim}.

In the above mentioned conditions, chiral symmetry will be sufficient to show that Landauer conductance is dominated by the leads, as occurs in $1d$ systems such as quantum wires, adiabatically connected to Fermi liquid reservoirs\cite{MS, OKM}. Then, following a line of reasoning similar to the one presented in refs. \onlinecite{OKM} and \onlinecite{univ}, a simple argument relying on general properties of the system will be given, in spite of the fact that the $2d$ sample contains complicated gapless modes, modeled as confined massless interacting Dirac fermions.

From a technical point of view, we will use the functional bosonization for $2d$ systems, where the current is mapped into a topological current, containing effective ``electric'' and ``magnetic'' fields for a vector potential $A_t$, $A_x$, $A_y$ (the bosonizing fields). In principle, a closed form for the dual bosonized action describing $2d$ Dirac fermions is only known in the large mass limit. However, on the leads, as the typical scale of one of the dimensions is small, we argue that results from $1d$ bosonization can be applied there. This amounts to approximating, on the leads, the $2d$ fermion determinant by considering only the contribution of fermion zero modes, and taking into account the decoupling of the other modes for small widths\cite{CesarTrinchero}. 

Another important ingredient in our derivation
will be the universal character of the bosonization rule for fermion currents in a general $2d$ system\cite{bos}.
In some sense, we will see that bosonization implements the idea of an electron wave guide, enabling 
the discussion of transport properties in terms of similar concepts associated with ``electromagnetic'' wave guides, such as geometric and material dispersion for the coupled modes. For instance, the decoupling of the zero modes on the quasi one-dimensional leads amounts to the suppression of the geometric dispersion, in wave guide language.

In section \S\ref{bosonization}, we describe the bosonization technique, stressing its physical meaning and showing the similarities and differences between $1d$ (\S\ref{1d}) and $2d$ systems (\S\ref{2d}). In particular, in \S\ref{2d} and \S\ref{2c}, we explain how to construct the bosonized action for the general system ($2d$ sample plus quasi one-dimensional leads), showing how the bosonization technique implements the idea of electron wave guide. In section \S\ref{quantization}, we derive the main result of this paper, showing the conductance quantization for a wide class of ultraclean systems with interactions. Finally, we discuss our results in \S\ref{discussions}.

\section{Chiral transport and bosonization \label{bosonization}}

\subsection{One dimensional massless fermions \label{1d}}

In this section we give a brief summary of one dimensional bosonization. Although the procedure is very well established\cite{coleman,mandelstam,luther,haldane}, we would like to emphasize 
the physical concepts involved that will lead us to new results, developed in the following sections. 

The description of a conserved charge in a $1d$ system,
is usually simplified by introducing a bosonic field $\phi$ such that,
\begin{equation}
\rho= \partial_x \phi
\makebox[.3in]{,}
j_x=-\partial_t \phi,
\label{eq1}
\end{equation}
which automatically leads to the continuity equation, 
\begin{equation}
\partial_t \rho+\partial_x j_x =0
\label{weq}
\end{equation}
for this reason, the $\phi$-field expressions are called a ``topological current''. 

For instance, we can consider a gapless $1d$ noninteracting fermionic mode $\psi$ characterized by a linear dispersion $E(p)=v p$. In quantum wires, this model can be used to represent those modes in the $2d$ Fermi liquid in the leads that couple to a quasi-one-dimensional quantum wire (sample), in this case $E$, and $p$ refer to energy and momentum measured with respect to a Fermi surface (points), at low excitation energies. In this case the field mode can be 
either right or left  moving ($\psi_R=\psi_R(x-vt)$ or $\psi_L=\psi_L(x+vt)$, respectively). Then, the current densities,
\begin{equation}
\rho=e(\rho_R +\rho_L)
\makebox[.5in]{,}
j_x=ev(\rho_R -\rho_L),
\label{densities}
\end{equation}
where 
$\rho_R=\psi_R^\ast\psi_R$ and $\rho_L=\psi_L^\ast\psi_L$, also satisfy, 
\begin{equation}
v^{-2}\, \partial_t j_x +\partial_x \rho=0,
\label{cheq}
\end{equation}
in addition to the continuity equation. Then, from eqs. (\ref{eq1}) and (\ref{cheq}), the field $\phi$ must obey the wave equation,
\begin{equation}
(v^{-2} \partial_t^2-\partial_x^2)\phi=0,
\label{fieq}
\end{equation}
and we can introduce an action,
\begin{equation}
S_0[\phi]=\int dt dx\, \frac{1}{2\alpha}\left[v^{-2}(\partial_t \phi)^2-(\partial_x\phi)^2\right], 
\label{S0}
\end{equation}
whose minimization leads to the wave equation (\ref{fieq}).

For the associated quantum theories, the quantum equivalence between a $1d$ massless Dirac field and a massless scalar field $\phi$, with the current mapping eq.\ (\ref{eq1}) is well known and is called bosonization \cite{coleman,mandelstam,luther,haldane}. For a general fermionic theory whose action is $K_F[\psi]$ having arbitrary short or long ranged interactions $I[\rho,j_x]$ only involving densities and currents, we have shown in ref. \onlinecite{bos} the following equivalence between partition functions, 
\begin{eqnarray}
Z&=&\int {\cal D}\psi {\cal D}\bar{\psi}\,  \exp iK_F[\psi] +i I[\rho,j_x]\nonumber \\
&=&\int {\cal D}\phi\, \exp iK_B[\phi] +i I[\partial_x \phi,-\partial_t \phi],
\label{1dmapping}
\end{eqnarray}
where $K_B$ can be computed as a ``transverse'' functional Fourier transformation 
of the partition function associated with $K_F$ (see ref. \onlinecite{univ}). Moreover, the computation of $K_B$ for a $1d$ massless Dirac field leads to
$K_B[\phi]=S_0[\phi]$, with $\alpha=(1/v)\, \frac{2e^2}{h}$.
In particular, for an external electric field, $I_e=\int d^2x\, \rho V$, in $\phi$-language we have a new term $I_e=\int d^2x\, (\partial_x \phi) V=\int d^2x\, \phi (-\partial_x V)$, so that the saddle point equation reads,
\begin{equation}
(v^{-2}\partial_t^2-\partial_x^2)\phi =\alpha E.
\label{anomaly}
\end{equation}
On the other hand, at the classical level, in addition to  the usual $U(1)$ symmetry $\psi_L \rightarrow e^{i\theta}\psi_L$, $\psi_R \rightarrow e^{i\theta}\psi_R$ leading to charge conservation, $1d$ massless Dirac fermions with current interactions posses a chiral symmetry,
$\psi_L \rightarrow e^{i\gamma}\psi_L$, $\psi_R \rightarrow e^{-i\gamma}\psi_R$. Because of the relative minus sign in the phases, the right and left mode contributions to the chiral density and current are,
\begin{equation}
\rho^A=e(-\rho_R +\rho_L )
\makebox[.5in]{,}
j^A_x=-ev(\rho_R +\rho_L),
\end{equation}
containing a change of sign with respect to the U(1) quantities in eq. (\ref{densities}), that is,
\begin{equation}
\rho^A=-j_x/v 
\makebox[.5in]{,}
j^A_x=-\rho v. 
\end{equation}
Because of this symmetry, in the classical system we have the conservation law,
\begin{equation}
\partial_t \rho^A+\partial_x j^A_x =0.
\end{equation}
 
However, if the action mapping in eq. (\ref{1dmapping})  is considered, which is needed for the quantum equivalence between the fermionic and bosonic partition functions, using the $\phi$-language in eq. (\ref{anomaly}), we get,  
\begin{equation}
\partial_t \rho^A+\partial_x j^A_x=v(v^{-2}\partial_t^2-\partial_x^2)\phi=(v\alpha) E.
\label{chiralanomaly}
\end{equation}
In other words, eq. (\ref{chiralanomaly}) represents the nonconservation of the chiral current at the quantum level. A classical symmetry that is not realized at the quantum level is called an anomaly. From a physical point of view, the anomaly represents the creation of particles and holes out of the vacuum or Fermi sea, when an external field is applied; of course, these concepts only exist in the quantum world. As it is well known, this effect fixes the value of $\alpha$ as follows. Consider an homogeneous electric field $E$ and an initial Fermi sea. From the equation of motion $\dot{p}=e E$ all right moving particles will gain a momentum $eEt$. If the system size is $L$, then the number of right moving particles created above the Fermi sea, at $p_F=0$, is the volume occupied in phase space, $L(eEt)$, divided by Planck's constant $h$, so that the density of right moving particles is $eEt/h$. 
This equals the density of left moving holes (or antiparticles) created, to conserve the system's charge equal to zero (with respect to the charge of the Fermi sea). This is measured by the quantum version of $\rho=e(\rho_R+\rho_L)$. Now, because of the relative minus sign in $\rho_A=e(-\rho_R+\rho_L)$, at the quantum level, it will measure the total density of particles (times the electric charge $e$): $\rho_A=(2e^2Et/h)$. Since the system is homogeneous, eq. (2) then gives,
\begin{equation}
\partial_t \rho_A+\partial_x j^A_x=2e^2 E/h,
\end{equation}
and the value,
\begin{equation}
\alpha=(1/v)\, \frac{2e^2}{h}
\end{equation} is obtained. Note that if $N$ dispersionless one-dimensional channels $\psi_i$, $i=1,..,N$ were considered, each one represented by a field $\phi_i$, the anomaly for the total chiral currents in eq. (\ref{chiralanomaly}) would be $(vN\alpha) E$, $\alpha=(1/v)\, \frac{2e^2}{h}$. Equivalently, if a single field $\phi$ were used to describe the total currents, then the value $\alpha=(1/v)\, N\frac{2e^2}{h}$, should be used for the associated field theory.

\subsection{Two dimensional four-component massless fermions \label{2d} confined to a strip}

Following the same reasoning of the previous section, let as consider a $2d$ system with a conserved charge, and the associated continuity equation,
\begin{equation}
\partial_t \rho+\partial_x j_x +\partial_y j_y=0.
\end{equation}
Formally, this equation is a ``three-divergence'' equal to zero, so that we can represent $\rho$, $j_x$ and $j_y$
as a ``curl'' or topological current,
\begin{eqnarray}
\rho&=& \partial_x A_y -\partial_y A_x\nonumber \\
j_x&=&\partial_y A_t -\partial_t A_y\nonumber \\
j_y&=&\partial_t A_x -\partial_x A_t,\nonumber \\
\label{3div}
\end{eqnarray}
which is identically conserved. 

Considering that $A_t$, $A_x$, $A_y$ and $A_t+\partial_t \chi$, $A_x+\partial_x \chi$, $A_y+\partial_y \chi$ represent the same charge density and current distribution, the $A$-fields must be gauge fields, physically equivalent when the above mentioned transformation is performed. It is useful to think about the charge density and currents as effective ``magnetic'' and ``electric fields''. Making the identifications, we have,
\begin{eqnarray}
\rho &=& B\nonumber \\
j_x&=&+E_y\nonumber \\
j_y&=&-E_x.\nonumber \\
\label{map}
\end{eqnarray}
Of course, $E$ and $B$ are not real electromagnetic fields, they are simply useful auxiliary fields to represent charge density and currents.
In particular, charge conservation now looks like a $2d$ Faraday-Lenz law,
\begin{equation}
\partial_x E_y -\partial_y E_x = -\partial_t B.
\label{F-L}
\end{equation}
Thus, charge variation in a given region is associated with nonzero current flux through the boundary, as changes in the magnetic flux piercing a surface are associated with an induced electric field, in bosonized language.

If gapless parity preserving fermions $\psi$ with a linear dispersion $E(\vec{p})=v|\vec{p}|$ (no ``material'' dispersion) are considered, we can guess that in $2d$ we will have to face a difficult problem because of ``geometrical'' dispersion. The fermion modes are  general combinations of waves $\exp i(\vec{k}.\vec{x}\pm \omega t)$ propagating with speed $v=\omega/|\vec{k}|$, satisfying the wave equation derived from Dirac's equation. But the charge density and currents are $\psi$-field  bilinears, not satisfying any simple equation, because of the continuum of possible directions given by $\vec{k}$ (``geometrical'' dispersion).  That is, if we try to write other equations defining $\rho$, and $\vec{j}$, or their effective electric and magnetic counterparts, they will certainly be highly nontrivial.

On the other hand, if gapless excitations were confined to a $2d$ strip, we expect that the effects of geometrical dispersion will decrease, as the strip width is reduced. Then, we could use the $\phi$-language representation of \S\ref{1d} for these modes. However, we would like to consider a general situation where the strip could be coupled adiabatically with an extended $2d$ region, where a continuum of gapless modes exist, with no definite direction of propagation, described by the gauge field $A$. For this reason, we will translate the dispersionless modes in the strip from $\phi$ to $A$-language.

If the strip is defined by a region of length $L$ along the $x$-axis, and width $W$, that is $y \in [-W/2,+W/2]$, the confinement condition of having no charge flux across the limits, that is $j_y=0$ at $y=\pm W/2$, amounts to $E_x=0$ at $y=\pm W/2$. In terms of the effective model this means that the the border acts as a perfect conductor or wave guide (see ref. \onlinecite{cesar1}). 

Assuming that the current distribution on the strip is all along the $x$-axis, that is $j_y=0$, while $\rho$ and $j_x$ are $y$-independent, we have,
\begin{equation}
\partial_x j_y -\partial_y j_x=0
\makebox[.5in]{,}
v^{-2}\,\partial_t j_y +\partial_y \rho=0,
\label{irrot}
\end{equation}
or in ``effective'' language,
\begin{equation}
\partial_x E_x +\partial_y E_y=0
\makebox[.5in]{,}
\partial_y B-v^{-2}\,\partial_t E_x=0.
\label{ME1}
\end{equation}
Now, using the condition (\ref{cheq}), for dispersionless $1d$ modes, we also have,
\begin{equation}
\partial_x B + v^{-2}\, \partial_t E_y=0.
\label{ME2}
\end{equation}
Eqs. (\ref{F-L}), (\ref{ME1}) and  (\ref{ME2}) are Maxwell's equations in $2d$, while the current distribution considered in the strip corresponds to an effective  dispersionless $TE_0$ mode coupled in a wave guide. These equations can be derived from $2d$ Maxwell's action,
\begin{equation}
S_0[A]=\int dt d^2x\, \frac{1}{2\beta}[v^{-2}(E_x^2+E_y^2)- B^2],
\label{Maxwell}
\end{equation}
with $E_x$, $E_y$ and $B$ defined through eqs. (\ref{3div}) and (\ref{map}).

We also note that in the radiation gauge $\partial_x A_x+\partial_y A_y=0$, $A_t=0$, and for the $TE_0$ mode we can use $A_x=0$, so that eq. (\ref{3div}) reads,
\begin{eqnarray}
\rho&=& +\partial_x A_y \nonumber \\
j_x&=&-\partial_t A_y.\nonumber \\
j_y&=&0.\nonumber \\
\end{eqnarray}
Then, considering that $\rho$ in $1d$ and $2d$ correspond to charge by unit length and area, respectively, while $j_x$ has units of current and current by transverse length, respectively, we can identify $W\rho$ and $Wj_x$ in $2d$ with one dimensional quantities, that is, we can identify $\phi\equiv W A_y$. We would like to underline that the dispersionless modes have a formal ``Lorentz'' invariance associated with boosts with speed parameter $v$. Under this boost $A_y$ is invariant, so that, in the quasi one-dimensional lead, its identification with the scalar $\phi$ is a natural one. 

Summarizing, if we evaluate Maxwell's action on the $TE_0$ mode, we obtain the identification $\beta=\alpha/W$, $\alpha=(1/v)\, N\frac{2e^2}{h}$, for the description of the total currents associated with $N$ dispersionless channels on a $2d$ strip be equivalent to $N$ $1d$ dispersionless quantum modes. 

\subsection{General chiral symmetric fermions confined to a 2d sheet} \label{2c}

For a general fermionic $2d$ model with interactions $I[\rho,j_x,j_y]$, only involving densities and currents,
the general quantum equivalence with the $A$-language description is given by,
\begin{eqnarray}
Z&=&\int {\cal D}\psi {\cal D}\bar{\psi}\,  \exp iK_F[\psi] +i I[\rho,j_x,j_y]\nonumber \\
&=&\int {\cal D}A\, \exp iK_B[A] +i I[B,E_y,-E_x],
\label{zbos2d}
\end{eqnarray}
where $K_F[\psi]$ is the free fermionic action.
The bosonizing action $K_B[A]$ is gauge invariant, and again is given by a ``transverse'' Fourier functional transform of the partition function associated with $K_F$\cite{bos}.

The low energy description of graphene can be written in terms of a four component spinor $\psi$, 
which can be organized as a pair $\psi_{K}$, $\psi_{-K}$, where the label is associated with excitations in the two opposite valleys centered at the corners of the Brillouin zone, with wave vector $\pm {\bf K}$ (Dirac points). The fields $\psi_{\pm K}$
are two-component Dirac spinors, each component is associated with the amplitude of the wave
function on the A and B sublattices of the honeycomb lattice. 

We underline that while the effective action for two-component massless fermions contains a Chern-Simons term\cite{point-splitting}, in the case of four component fermions, each element of the pair must contribute with an opposite sign, in order to conserve parity. Then, as the model for graphene is based on a four-component $\psi$, $K_B$ contains no Chern-Simons term, and being gauge invariant it must be necessarily of the form, $K_B=K_B[B,E_x,E_y]$. 

In general, if $K_F$ were a gapped (massive) theory, it would be possible to obtain a low energy expansion for $K_B$. However, in graphene the excitations are associated with massless Dirac fermions, and in an extended $2d$ region $K_B[A]$ is an unknown complicated functional, as there is no parameter to organize a perturbative expansion. 

On the other hand, from the previous discussion, on a thin $2d$ strip, $K_B[A]$ is expected to have the simple Maxwell form of eq. (\ref{Maxwell}), as the problem can be considered as quasi one-dimensional. To be more precise, a massless theory confined to a thin strip will have zero modes plus other modes that decouple in the small $W$
limit (see ref. \onlinecite{CesarTrinchero}). Thus, the partition function for $K_F$ on the strip is expected to be dominated by these zero modes; these are precisely the relevant modes to discuss transport in ultraclean graphene strips (see ref. \onlinecite{CN2006}).

Then, in the more general case, where a two dimensional system is adiabatically coupled to quasi one-dimensional leads, the mapping (\ref{zbos2d}) can be used, if corrections to the simple Maxwell form in the leads are 
incorporated in the bulk expression for $K_B$.

\section{Conductance of a general graphene strip/sample/strip system} \label{quantization}

From the previous discussion, if an extended $2d$ region with ultraclean graphene (gapless Dirac fermions) is adiabatically coupled to two quasi one-dimensional strips, formed with the same material, the system's bosonized action is expected to have the form,
\begin{equation} 
S_B=K_B+I
\makebox[.5in]{,}
K_B=S_0+R,
\end{equation}
where $S_0$ is given by eq. (\ref{Maxwell}), $R$ represents deviations from the simple Maxwell form valid on the leads, localized in the sample's bulk. The $I$-term represents charge density and current interactions also localized at the bulk In the present bosonized language, these interactions and the deviations $R$ play a similar role. In addition, the
``electric'' field $E_x$, $E_y$ must be orthogonal to the sample's boundary\cite{cesar1} (no charge flux across the boundaries).

The important point is that because of the adiabatic condition, $K_B$ is defined by a single functional of $B$, $E_x$ and $E_y$, whose form depends on the system's (lead/sample/lead) point under consideration. Although the corrections in the bulk are very difficult to compute, we will see that the above mentioned general structure for $K_B$ is all we need to derive the conductance quantization. 

To discuss transport, we include in eq. (\ref{zbos2d}) a simple coupling with external electric and magnetic fields $\{ {\cal A}_0,{\cal A}_x,{\cal A}_y\}$ to probe the system, that is, we consider the replacement, 
\begin{equation}
I\rightarrow I+I_e
\makebox[.5in]{,}
I_e=\int d^3x\,(-\rho {\cal A}_0+ j_x {\cal A}_x +j_y {\cal A}_y),
\end{equation}
or in $A$-language,
\begin{equation}
I\rightarrow I+ \int d^3x\,( E_y {\cal A}_x -E_x {\cal A}_y-B {\cal A}_0).
\end{equation}

The associated saddle point equations are obtained by taking functional derivatives with respect to $A_t$, $A_x$ and $A_y$, respectively,
\begin{eqnarray}
&&\partial_x \frac{\delta S_B}{\delta E_x}+\partial_y \frac{\delta S_B}{\delta E_y}={\cal B}\nonumber \\
&&\partial_y \frac{\delta S_B}{\delta B}+\partial_t \frac{\delta S_B}{\delta E_x}=-{\cal E}_y\nonumber \\
&&\partial_x \frac{\delta S_B}{\delta B}-\partial_t \frac{\delta S_B}{\delta E_y}=-{\cal E}_x
\label{saddlepoint}
\end{eqnarray}
where we have defined the system's bosonized action,
\begin{equation}
S_B=K_B + I.
\label{SB}
\end{equation}

Now, we can follow a  reasoning very similar to the one of Maslov and Stone\cite{MS} used to derive the universal behavior of conductance in quantum wires (see also ref. \onlinecite{OKM}).

Let us consider an electric field that is switched on to attain a stationary value ${\cal E}_x(x,y)$, ${\cal E}_y(x,y)$. At late times a uniform current $I$ is expected to be settled in the system, so that on the strips we can consider an ansatz $A_t=0$, $A_x=0$ and 
$A_y=f(x)-kt$, $I/W=j_x=-\partial_t A_y=k$, so that $I=k W$. 

Because of causality, $A_y$ on the left (right) lead must be given by a perturbation propagating to the left (right), with speed $v$. Then, considering symmetry under $x$-reflection (to simplify the argument), $f(x)$ must have the form $-(k/v)x$ on the left lead and $+(k/v)x$ on the right lead. This ansatz can be extended to a solution on the whole system, and considering the last two equations in (\ref{saddlepoint}), we have,
\begin{eqnarray}
&&\partial_y \left[-(1/\beta)B+ \frac{\delta (R+I)}{\delta B}\right]=-{\cal E}_y\nonumber \\
&&\partial_x \left[-(1/\beta)B+\frac{\delta (R+I)}{\delta B}\right]=-{\cal E}_x.
\end{eqnarray}
Integrating on a curve ${\cal C}$ going from the left to the right lead,
\begin{equation}
-(1/\beta)[B|_{right}-B|_{left}]=-\int_{\cal C} (dx {\cal E}_x + dy {\cal E}_y)=\Delta V,
\end{equation}
and using eqs. (\ref{3div}) and (\ref{map}), and the ansatz above,
\begin{eqnarray}
B|_{right}&=&(\partial_x A_y -\partial_y A_x)_{right}=+\frac{k}{v}\nonumber \\
B|_{left}&=&(\partial_x A_y -\partial_y A_x)_{left}=-\frac{k}{v}.
\end{eqnarray}
Therefore, we obtain,
\begin{equation}
I=-(Wv\beta/2) \Delta V= -N \frac{e^2}{h}\Delta V,
\label{conductance}
\end{equation}
where $N$ is the number of zero fermion modes on the strips.

\section{Summary and discussion \label{discussions}}

We have analyzed the problem of linear transport in an arbitrarily shaped ultraclean graphene sheet, adiabatically connected to quasi one-dimensional leads, formed with the same material. This means that a smooth widening 
of the strips is permitted at the sample's bulk, where effective interactions are localized, which depend on the total charge density and currents associated with both Dirac points. Under this circumstances, we have shown that the conductance is quantized, as it is dominated by the ultraclean quasi one-dimensional leads. 

The situation is similar to what happens in quantum wires, where conductance is quantized because of the dominance of the Fermi liquid reservoirs.

There is an interesting realization of the scenario discussed here, recently studied in the literature, quite relevant when one tries to compare models with actual experiments.
In ref. \onlinecite{Kim}, Castro Neto and Kim analyzed the effect of the curvature of the (otherwise plane) carbon sheet on the electronic structure of the sample. 
They showed that the net effect of smooth curvature fluctuations is that fermions become minimally coupled with an effective ``elastic'' gauge field.
Permanent deformations could eventually break time reversal symmetry at one point. However the effect on the other Dirac point $-K$ restores this symmetry. In our formalism, that means that the bosonized action $K_B$ has no  induced Chern-Simons term, due to the cancellation between both species of fermions. On the other hand, the elastic field couples symmetrically the two Dirac points, meaning that  smooth fluctuations of this gauge field will  induce total charge and current interactions of the class considered in this paper, so that our quantization argument can be applied in this case. 

In general, our argument works whenever the chiral symmetry displayed by $2d$ four-component massless fermions is preserved, so that the obtained perfect conductance can be associated with effective gapless excitations defined along the whole system. Bosonization gives a nice interpretation for this phenomenon, looking to the sample as an ``electromagnetic'' wave guide, without backscattering, guiding the $TE_0$ modes at the leads. 

\begin{acknowledgments}
The Conselho Nacional de Desenvolvimento Cient\'{\i}fico e Tecnol\'{o}gico (CNPq) and the Funda{\c {c}}{\~{a}}o de Amparo {\`{a}} Pesquisa do Estado do Rio de Janeiro (FAPERJ) are acknowledged for the financial support.
\end{acknowledgments}

\end{document}